\documentclass[aj,twocolumn]{aastex62}

\usepackage{placeins}
\usepackage{comment}
\usepackage{graphicx}
\newcommand{\thisstar}{HIP 94235}

\usepackage{wrapfig}
\usepackage{lineno}
\submitjournal{AJ}

\shorttitle{HIP94235b Helium}
\shortauthors{Morrissey et al.}
\begin{document}

\title{Searching for helium escape and a low density atmosphere around the 120 Myr old sub-Neptune HIP94235b using CRIRES+}

\correspondingauthor{Ava Morrissey}
\email{ava.morrissey@unisq.edu.au}

\author[0009-0000-3527-8860]{Ava Morrissey} 
\affil{University of Southern Queensland, Centre for Astrophysics, West Street, Toowoomba, QLD 4350, Australia}

\author[0000-0002-4891-3517]{George Zhou}
\affil{University of Southern Queensland, Centre for Astrophysics, West Street, Toowoomba, QLD 4350, Australia}

\author[0000-0003-0918-7484]{Chelsea X. Huang}
\affil{University of Southern Queensland, Centre for Astrophysics, West Street, Toowoomba, QLD 4350, Australia}

\author[0000-0001-7294-5386]{Duncan Wright}
\affil{University of Southern Queensland, Centre for Astrophysics, West Street, Toowoomba, QLD 4350, Australia}

\author[0000-0002-9308-2353]{Neale Gibson}
\affil{School of Physics, Trinity College Dublin, University of Dublin, Dublin 2, Ireland}

\author[0000-0003-1337-723X]{Keighley E. Rockcliffe}
\affil{Exoplanets and Stellar Astrophysics Laboratory, NASA Goddard Space Flight Center, Greenbelt, MD, USA}

\author[0000-0003-4150-841X]{Elisabeth R. Newton}
\affil{Department of Physics and Astronomy, Dartmouth College, Hanover, NH 03755, USA}

\author[0000-0002-4207-6615]{James Kirk}
\affil{Department of Physics, Imperial College London, Prince Consort Road, SW7 2AZ, London, UK}

\author[0000-0001-6023-1335]{Daniel Bayliss}
\affil{Department of Physics, University of Warwick, Gibbet Hill Road, Coventry CV4 7AL, UK}
\affil{Centre for Exoplanets and Habitability, University of Warwick, Gibbet Hill Road, Coventry CV4 7AL, UK}

% %% Note that the \and command from previous versions of AASTeX is now
% %% depreciated in this version as it is no longer necessary. AASTeX 
% %% automatically takes care of all commas and "and"s between authors names.

% %% AASTeX 6.2 has the new \collaboration and \nocollaboration commands to
% %% provide the collaboration status of a group of authors. These commands 
% %% can be used either before or after the list of corresponding authors. The
% %% argument for \collaboration is the collaboration identifier. Authors are
% %% encouraged to surround collaboration identifiers with ()s. The 
% %% \nocollaboration command takes no argument and exists to indicate that
% %% the nearby authors are not part of surrounding collaborations.

% %% Mark off the abstract in the ``abstract'' environment. 

\begin{abstract}
Atmospheric mass loss is thought to induce the bimodality in the small planet population as we observe it today. Observationally, active mass loss can be traced by excess absorption in spectral lines of lighter species, such as the hydrogen Ly-$\alpha$ line and the metastable helium triplet. We search for helium escape from the young (120Myr old) sub-Neptune (3$R_\oplus$) HIP94235b. We obtained two transit observations of HIP94235b using the CRyogenic InfraRed Echelle Spectrograph (CRIRES+) on the Very Large Telescope (VLT). We find no evidence for escaping helium across both visits, allowing us to place a mass loss rate upper limit of $10^{11}$gs$^{-1}$, based on 1D Parker wind models. Additionally, we search for molecular spectral features in the planet's transmission spectrum, and cross-correlate our observations with high-resolution template spectra for H$_2$O, the dominating molecule in the Y-band. We detect no significant absorption. We demonstrate that some atmosphere models at $10\times$ solar metallicity would have been retrievable if present. Through the null detection of neutral hydrogen and helium escape, we conclude the atmosphere of HIP94235b likely lacks a large hydrogen-helium envelope. This is consistent with the expectation of small planet photoevaporation models, which suggest most planets lose their primordial hydrogen-helium envelopes within 100Myr of evolution. 
\end{abstract}

\keywords{%%
    planetary systems ---
    stars: individual (\thisstar)
    techniques: spectroscopic, photometric
    }


% #####################################################################
%% Introduction
%% will start to completely rewrite this section after methods
\section{Introduction}
\label{sec:introduction}
% Atmospheric escape is the fundamental process that is responsible for shaping the small planet population into their two distinct groups; the smaller super Earths and the larger sub-Neptunes. While atmospheric escape is a process that affects the entire exoplanet population, it is a pivotal mechanism for these smaller (x R, y M) planets, as it can directly alter the trajectory of their evolution.\\ 
Transit observations that span the metastable helium triplet of lines wavelength region allow for tracing of neutral helium escape from planets \citep{Spake_2018, Oklop_2018}. Although the first species used to trace atmospheric escape was the hydrogen Ly-$\alpha$ line \citep[e.g.][]{vidalmadjar2003osirishd209458b}, the metastable helium triplet of lines of lines offers a unique perspective with different advantages. The helium infrared triplet is not affected by line core absorption from the interstellar medium (ISM). This expands the population of available planets suitable for observations beyond the $\sim 50$\,pc limit imposed on equivalent Ly-$\alpha$ observations. The lack of ISM absorption also allows any planetary absorption to be measured over all velocity spaces, in contrast to being restricted to highly blue or red shifted outflows in the case of Ly-$\alpha$. Helium, however, is most effective as a tracer of atmospheric escape around late type-stars (in particular K-type) due to their ionising potential, presenting limitations for observations of other stellar types \citep{Seager_Sasselov_2000, Turner_2016, Oklop_2019}.

Detecting ongoing escape from young sub-Neptunes is one pathway to understanding the formation and evolution history of small close-in planets surrounding the radius gap. These small planets may have formed in gas-rich environments, and have low mean molecular weight gaseous envelopes surrounding rocky cores \citep[e.g.][]{2014ApJ...797...95L,2016ApJ...825...29G}. Alternatively, small planets may have formed in the water-rich parts of the protoplanet disk, and are composed of water-rich exteriors surrounding Earth-like cores \citep[e.g.][]{2019PNAS..116.9723Z}. Importantly, gaseous envelopes are most susceptible to undergo runaway mass loss, while heavier water-rich envelopes undergo less significant changes in the early lives of these planets \citep[e.g.][]{2012ApJ...761...59L,2025ApJ...979...79R}.

Detections of helium escape have been reported for mature aged Neptune-sized planets previously, including HAT-P-11b \citep{Allart_2018, 2018ApJ...868L..34M,2024A&A...686A..83G} and GJ3470b \citep{Palle_2020, 2020ApJ...894...97N,2024A&A...686A..83G}. These planets orbit gigayear old systems, and are unlikely to undergo the runaway mass loss experienced by some Neptune-sized planets that would lead to large radial evolution over the 100\,Myr timescale \citep{Owen_2017,Ginzburg_2018}. Of the young planets surveyed so far, most have yielded non-detections of escaping neutral helium, including planets in the V1298 Tau \citep{2024AJ....168..102A,2024A&A...689A.179O}, TOI-1807 \citep{2023MNRAS.518.3777G,2024A&A...689A.179O}, and K2-100 \citep{2020MNRAS.495..650G,2024AJ....168..102A} systems. Tentative detections were reported for young ($<$ 1Gyr) sub-Neptunes in HD63433 c ($\sim$ 400 Myr) \citep{2022AJ....163...68Z}, TOI-560b (480-750 Myr), TOI-1430.01 (165 $\pm$ 30 Myr),  TOI-2076b (204 $\pm$ 50) and TOI-1683.01 (500 $\pm$ 150) \citep{Zhang_2023}, though follow-up observations have failed to confirm some of these detections \citep{2023MNRAS.518.3777G,2024AJ....168..102A}. 

% Previous works have surveyed for excess helium absorption in populations of planets, finding correlations with stellar spectral type and high energy irradiation \citep[e.g.][]{2024A&A...686A..83G}. 

The presence of a low density gas-rich envelope should also be linked to the bulk density and atmospheric metallicity of the planets. JWST and HST observations of young sub-Neptune progenitors such as V1298 Tau b, V1298Tau c \citep{2024NatAs...8..899B,2024A&A...692A.198B} and HIP67522b \citep{2024AJ....168..297T} ($\leq$20 Myr) have Neptune-like masses and sub-solar metallicities. Older sub-Neptunes ($>$1 Gyr), including GJ1214b \citep{2023ApJ...951...96G}, GJ9827d \citep{2024arXiv241003527P}, TOI-836c \citep{2024AJ....168...77W}, GJ3090b \citep{2025arXiv250316608P} and GJ436b \citep{2025arXiv250217418M}, have been shown to have significantly enhanced atmospheric metallicities ($\geq$100x solar).

%To understand where HIP94235b fits within the broader population of sub-Neptunes, we explore a discrete set of atmospheric metallicities (1x, 10x and 100xsolar) and masses (5, 10M$_\oplus$) in Section~\ref{sec:injection_recovery} of this paper. Young sub-Neptune progenitors such as V1298 Tau b, V1298Tau c \citep{2024NatAs...8..899B,2024A&A...692A.198B} and HIP67522b \citep{2024AJ....168..297T} ($\leq$20 Myr) exhibit sub-solar metallicities, whereas older sub-Neptunes ($>$1 Gyr), including GJ1214b \citep{2023ApJ...951...96G}, GJ9827d \citep{2024arXiv241003527P}, TOI-836c \citep{2024AJ....168...77W}, GJ3090b \citep{2025arXiv250316608P} and GJ436b \citep{2025arXiv250217418M}, show significantly enhanced atmospheric metallicities ($\geq$100x solar). In the second half of this paper, we run injection-recovery analysis tests to determine which combinations of planetary mass and metallicity can be ruled out, further constraining HIP94235b's place within the sub-Neptune evolutionary sequence.

HIP 943235 b is a 120$\pm$50\,Myr old sub-Neptune orbiting a G-type host ($T_{\rm eff}$ = 5991 K, 1.094 M$_\odot$, 1.08 R$_\odot$) on a 7.7 day period \citep{Zhou_2022}. The continuum transit depth of HIP94235b is 600ppm, or 0.06$\%$. \citet{Zhou_2022} estimated the age of HIP94235 based on its association with the AB Doradus moving group. Additionally, the star's stellar rotation rate, lithium abundance and x-ray emission intensity are all consistent with this age estimate. At 120\,Myr old, HIP94235b offers a unique time frame to search for escape around a star whose intense XUV bombardment of its orbiting planets is exponentially declining. The relatively small size (R = 3.00$^{+0.32}_{-0.28}$R$_\oplus$) and age of HIP94235b present a rare opportunity to investigate the evolutionary transition of sub-Neptunes to super-Earths. The small size of HIP94235b in comparison to other young, pre-main sequence Neptunes (e.g. K2-33b, and those in the V1298 Tau system) may suggest it is more akin to the class of super-Earths around more mature stars, which makes it an important object for understanding the effects of atmospheric escape on the small planet population.

\citet{Morrissey2024} searched for traces of escaping neutral hydrogen from HIP94235b. Though a null detection was reported from two Lyman-$\alpha$ transit observations, significant interstellar extinction of the line over a distance of 58\,pc prevented strong constraints from being placed on escape. In this paper, we present two transit observations of HIP94235b with the infrared high resolution spectrograph CRIRES+ (the recently upgraded CRyogenic high-resolution InfraRed Echelle Spectrograph) to search for traces of helium escape from the sub-Neptune. These helium observations are unobstructed by interstellar absorption, and any escaping tails are affected by the stellar ionizing irradiation differently to that from neutral hydrogen, and as such provide a new dimension to constraining the atmospheric composition and evolution of this sub-Neptune system. We also search for transmission spectroscopic signatures from water vapour in the atmosphere of the planet. Detections of such signatures are expected if the planet hosts a low metallicity, low mean molecular weight atmosphere that may still remain post formation.

\section{Observations and spectral extraction}
\label{sec:obs}

\subsection{Observations}

\begin{figure*}
    \centering
    \includegraphics[width=0.8\linewidth]{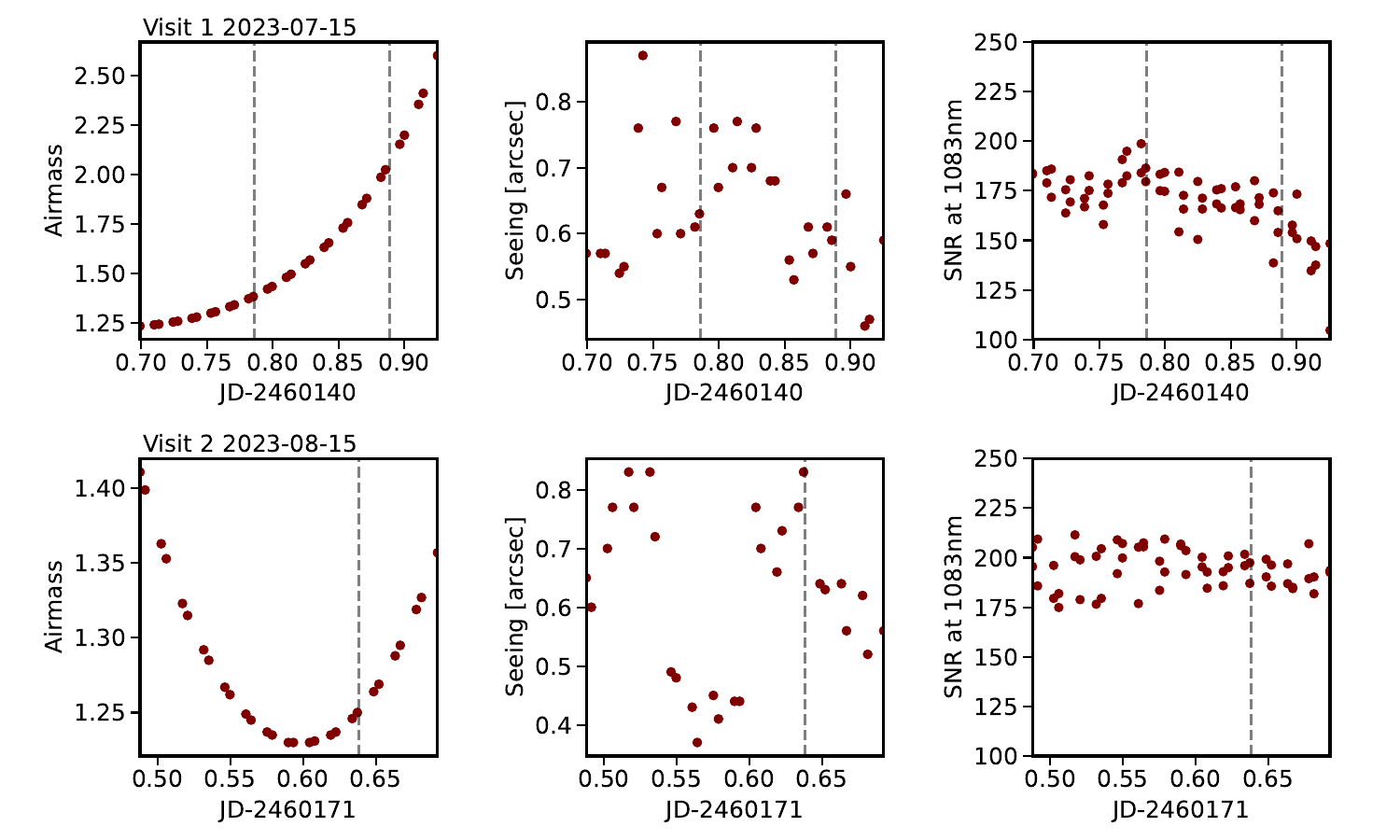}
    \caption{Environmental variations over the course of both transit observations. Included are variations in airmass, seeing, and per pixel signal to noise over the helium triplet at 1083\,nm over the course both nights. The vertical lines across all plots represent times of ingress and egress for both visits, respectively.}
    \label{fig:environmental}
\end{figure*}

We obtained two visits of HIP94235b with CRIRES+ on the VLT in La Paranal, Chilè \citep{Dorn_2023}. For both of our visits on July 14th and August 15th 2023, CRIRES+ observed HIP94235b for one half of each night, spanning 5.5 hours total observing time each visit. Observations in Visit 1 began two hours pre-ingress, and continued to observe the target through to 45 minutes post-egress. Observations in Visit 2 began 3 hours and 20 minutes pre-ingress, obtaining a solid pre-transit baseline. For both visits, CRIRES+ was operating at maximum resolution of $R=92000$ through the use of the 0.2" slit. HIP94235 has a brightness of V = 8.31$\pm$ 0.03 in the optical \citep{Henden_2016} and K = 6.881$\pm$ 0.027 in the infrared \citep{Skrutskie_2006}. The slit is placed at 90$^\circ$ with respect to sky to minimize the effects of atmospheric dispersion over the slit. A nodding pattern of ABBA was adopted for both visits to allow for efficient background removal. Visit 1 is broken up into 62 observations: 31 from nodding position A and 31 from B. Within each observation, there are nine spectral orders, corresponding to different wavelength regions across the spectrograph. Similarly, 28 observations in each nodding position were obtained in visit 2. Visit 2 observations benefited from adaptive optics corrections, with the target star acting as the natural guide star. Wavelength calibration is provided via day-time Neon/Krypton lamp exposures, as is standard to support CRIRES observations. Table~\ref{tab:obs} outlines a summary of the observations, and environmental variations over the course of the two nights are presented in Figure~\ref{fig:environmental}.\\

% \begin{figure*}[t]
%     \centering
%     \includegraphics[width=0.8\linewidth]{smallplots.png} 
%     \caption{Airmass (left), seeing (middle) and the flux over the helium line (right) plots for the course of the observations for both visit 1 (top) and visit 2 (bottom).}
%     \label{fig:smallplots}
% \end{figure*}

\begin{table}
    \centering
    \begin{tabular}{p{2.3cm}p{3.5cm}}
        \hline
        Target & HIP94235b\\
        Programme ID & 111.24TH.001(2023-07-15),
        111.24TH.002(2023-08-15)\\
        Instrument & CRIRES+\\
        Filter & Y1029\\
        Wavelength Coverage & 950-1120 nm \\
        Exp. Time & 1x900s\\
        AO & No (.001), Yes (.002)\\
        Airmass & See Fig.~\ref{fig:environmental}\\
        Seeing & See Fig.~\ref{fig:environmental}\\
        Slit & 0.2"\\
        \hline
    \end{tabular}
    \caption{Overview of both visits of HIP94235b using CRIRES+. Visit 1 is denoted as .001 and Visit 2 is denoted as .002. The exposure time of 1x900s represents NDITxDIT, which is the number of detector integrations x detector integration time.}
    \label{tab:obs}
\end{table}

\subsection{Spectral Extraction and calibration}
\begin{figure*}[t]
    \centering
    \includegraphics[width=0.9\linewidth]{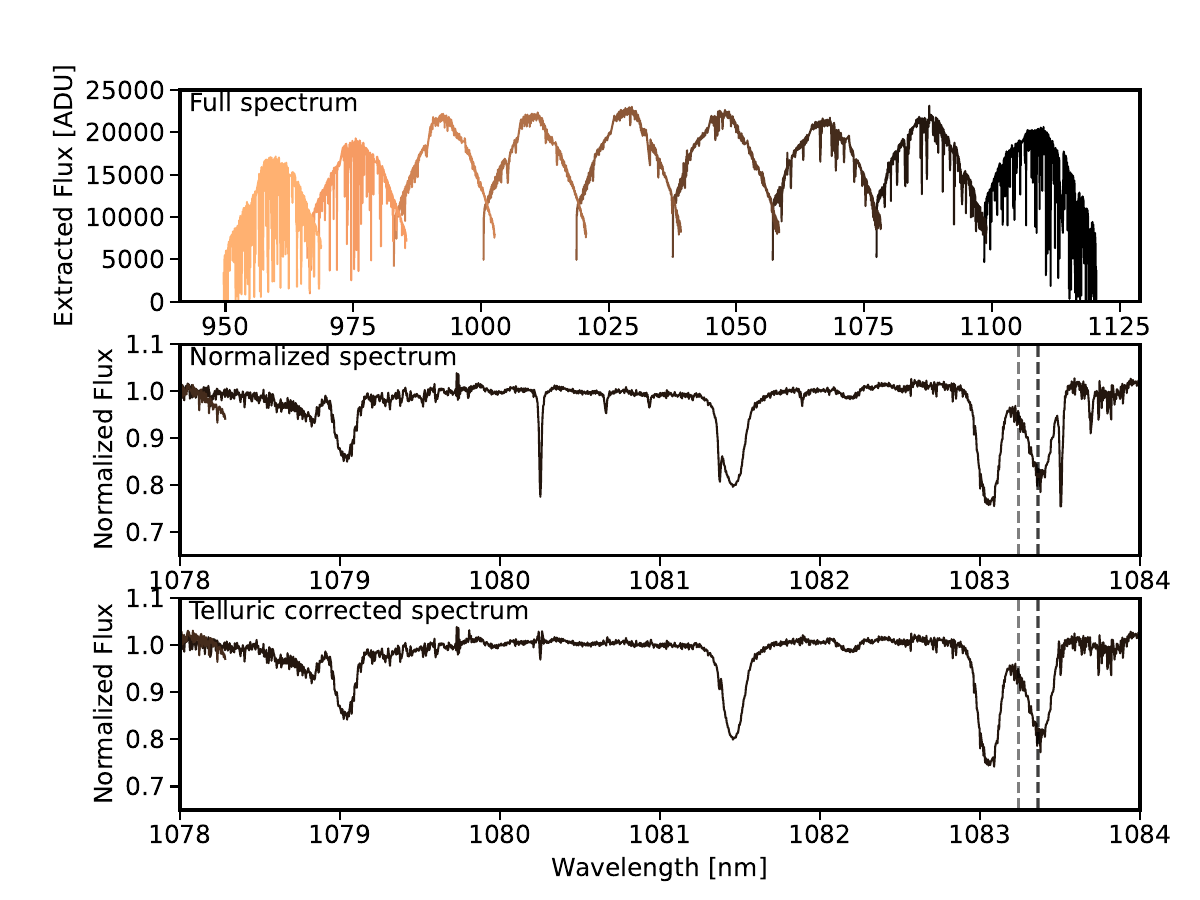} 
    \caption{One example CRIRES+ spectrum of HIP94235b obtained during visit 1. \textbf{Top} Extracted spectrum from all nine spectral orders, arranged in descending order from left to right, with order 9 on the far left and order 1 on the far right. \textbf{Middle} Normalized stellar spectrum over the helium infrared triplet at 1083\,nm. The wavelengths of the helium infrared triplet are marked by the vertical lines. \textbf{Bottom}  Telluric corrected portion of the spectrum over the helium infrared triplet region. }
    \label{fig:spec_v1}
\end{figure*}

We made use of \textsc{pycrires}, a data reduction pipeline for VLT/CRIRES+ \citep{pycrires}, a python wrapper for CRIRES+ pipelines of \textsc{EsoRex}, for calibration and spectral extraction. Figure~\ref{fig:spec_v1} shows one example extracted spectrum for Visit 1. Each order in the upper panel shows the raw spectra prior to normalization and telluric correction. The edges of each order have been trimmed by 50 pixels on each side. The lower panel depicts order 2, spanning the 1077 - 1084\,nm wavelength range, surrounding the 1083 nm metastable helium triplet of lines. The spectrum in the lower panel has been corrected for telluric absorption, following processes described in Section~\ref{sec:telluric}. 

\subsection{Telluric correction}\label{sec:telluric}
We make use of the \textsc{telfit} script for telluric correction of our observed spectra \citep{1992JGR....9715761C, 2014AJ....148...53G}. Telfit models the temperature, pressure, telescope zenith angle, and abundances of the telluric atmosphere, and wavelength offsets of the input dataset. This is performed iteratively while also accounting for the stellar spectrum via the PHOENIX model library \citep{1999ApJ...512..377H}. 

Examples of the telluric corrected spectra are shown in Figure~\ref{fig:spec_v1}. Strong and saturated telluric lines remain poorly subtracted after this procedure, and are masked out manually for further analysis. Orders 1, 8 and 9 are heavily contaminated by tellurics, so we discarded these orders for further analysis.

\section{Searching for neutral helium escape}\label{sec:helium}

\subsection{Removal of stellar signal}
\begin{figure}[h!]
    \centering    \includegraphics[width=0.9\linewidth]{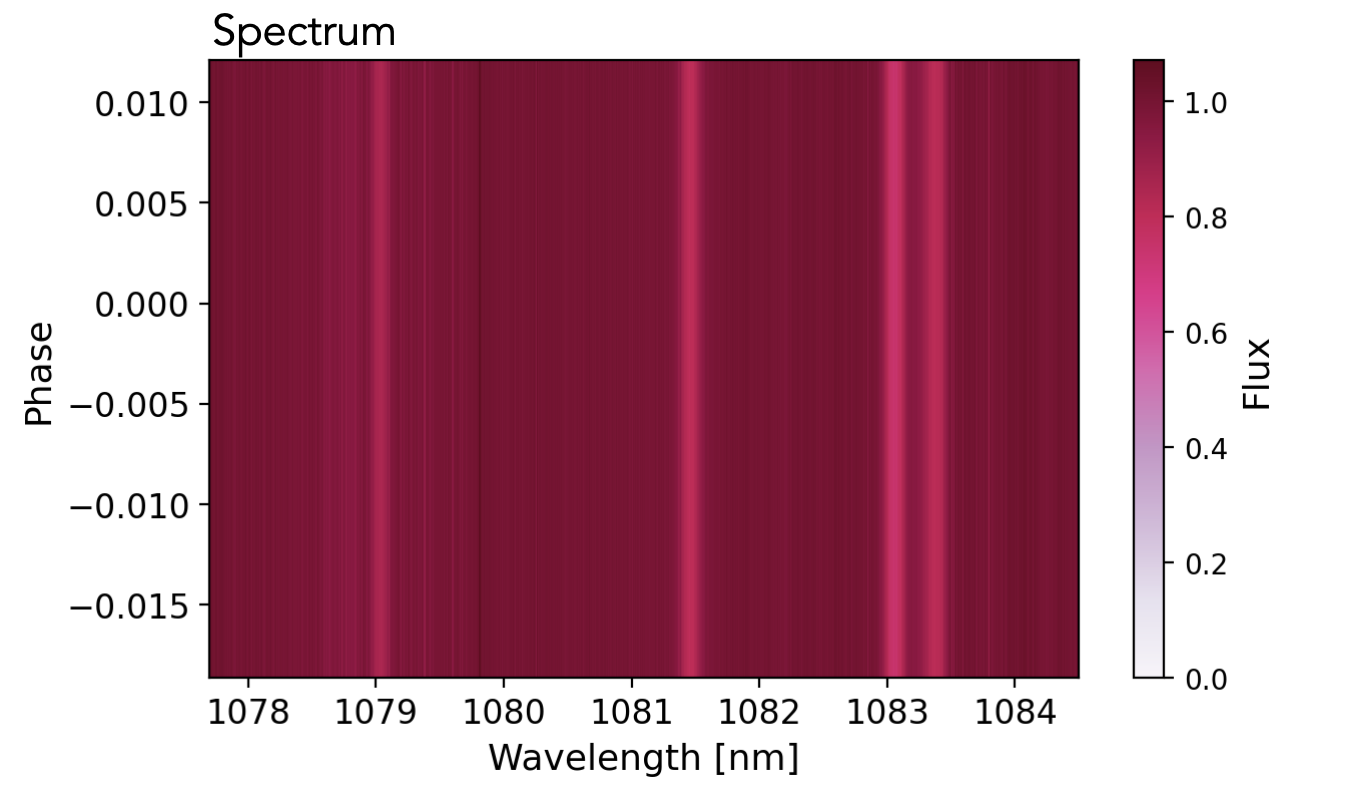} 
    \includegraphics[width=0.9\linewidth]{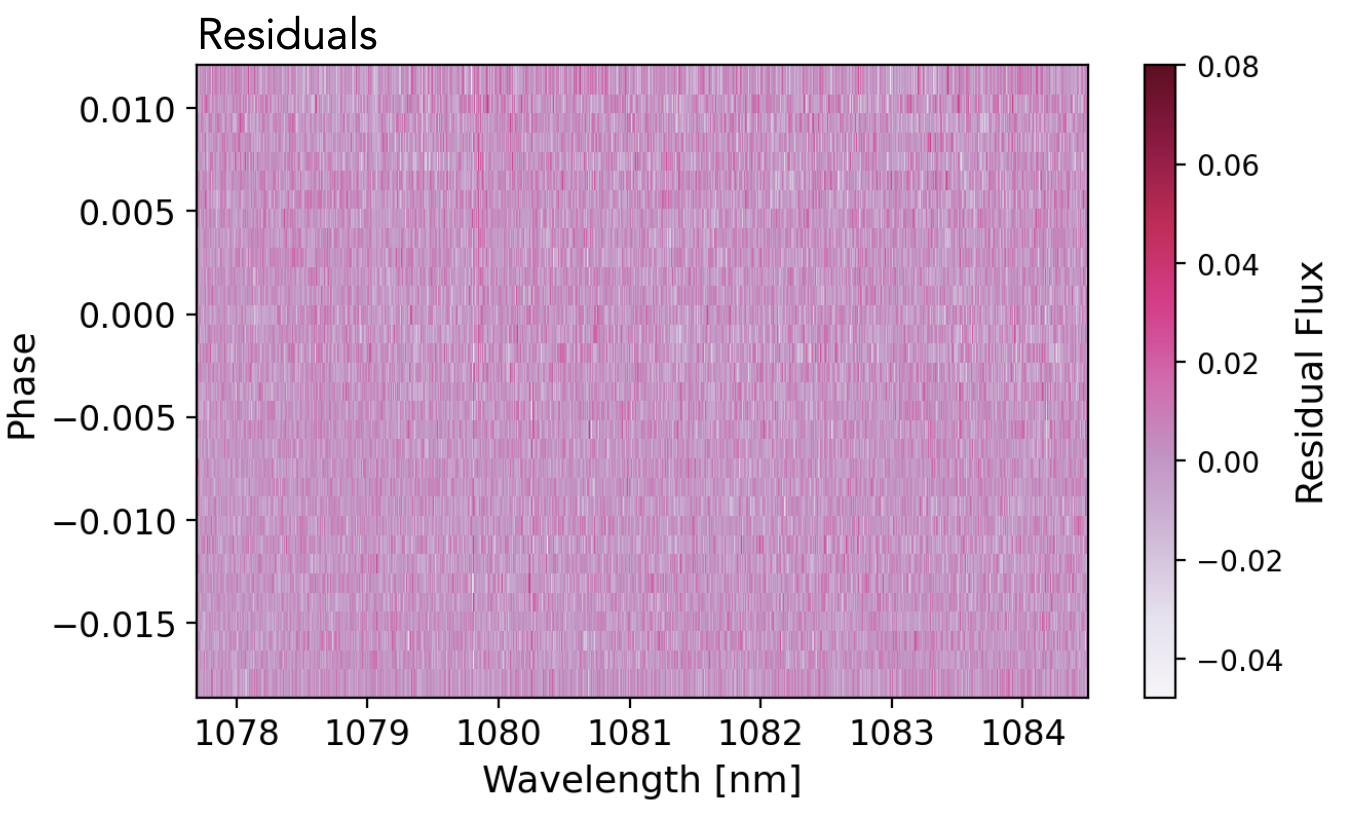}
    \caption{We make use of a median out-of-transit stellar spectrum to remove the stellar spectral signal over the helium line triplets range. Temporal variations of the stellar spectrum are shown. The median removal is performed on a per-visit, per nodding position basis, as each visit and nodding position has its own associated systematics. \textbf{Top:} Observations over the first CRIRES+ visit, from nodding position A, prior to median division. Stellar absorption features appear as deep vertical stripes that are constant in velocity within the stellar rest frame. \textbf{Bottom:} The spectral residuals after applying median division, demonstrating the removal of stellar spectral features while preserving any potential planetary signal.}
    \label{fig:medsubt}
\end{figure}

Post-telluric calibration, we corrected for stellar contamination via a median division of an observed master template. We computed the median stellar spectrum from out-of-transit exposures and divided it from each individual exposure, removing quasi-static stellar features while preserving time-dependent signals. We applied this method to all orders for each observation across both nights, producing residual maps for further analysis. Fig~\ref{fig:medsubt} shows the effect of this approach on visit 1 of the CRIRES+ data. The upper panel shows the observed spectra before median removal, while the lower panel presents the residuals after the stellar removal. A discussion of \textsc{SysRem} and why it is not suitable to this dataset and this planet is provided in Section~\ref{sec:injection_recovery}.
% Post-telluric calibration, we corrected for stellar contamination via the \textsc{SysRem} algorithm \citep{Tamuz_2005}. \textsc{SysRem} is a modification of the principal component analysis technique that accounts for non-uniform weightings of the observations. \textsc{SysRem} removes stationary information from time series observations, such as the stagnant stellar signal, while leaving the velocity shifted planetary signal for further analysis. We applied \textsc{SysRem} to all orders for each observation across both nights, giving us a residual map of any remaining planetary signal for each order, containing the information from each time-dependent observation. Figure~\ref{fig:medsubt} shows the effect of \textsc{SysRem} on visit 1 of the CRIRES+ data. It begins by creating a model spectrum based on the observed data, shown in the upper panel. The bottom panel represents the residuals after 2 iterations of \textsc{SysRem}. 

\subsection{Search for excess helium absorption}
\begin{figure}[h!]
    \centering    \includegraphics[width=0.9\linewidth]{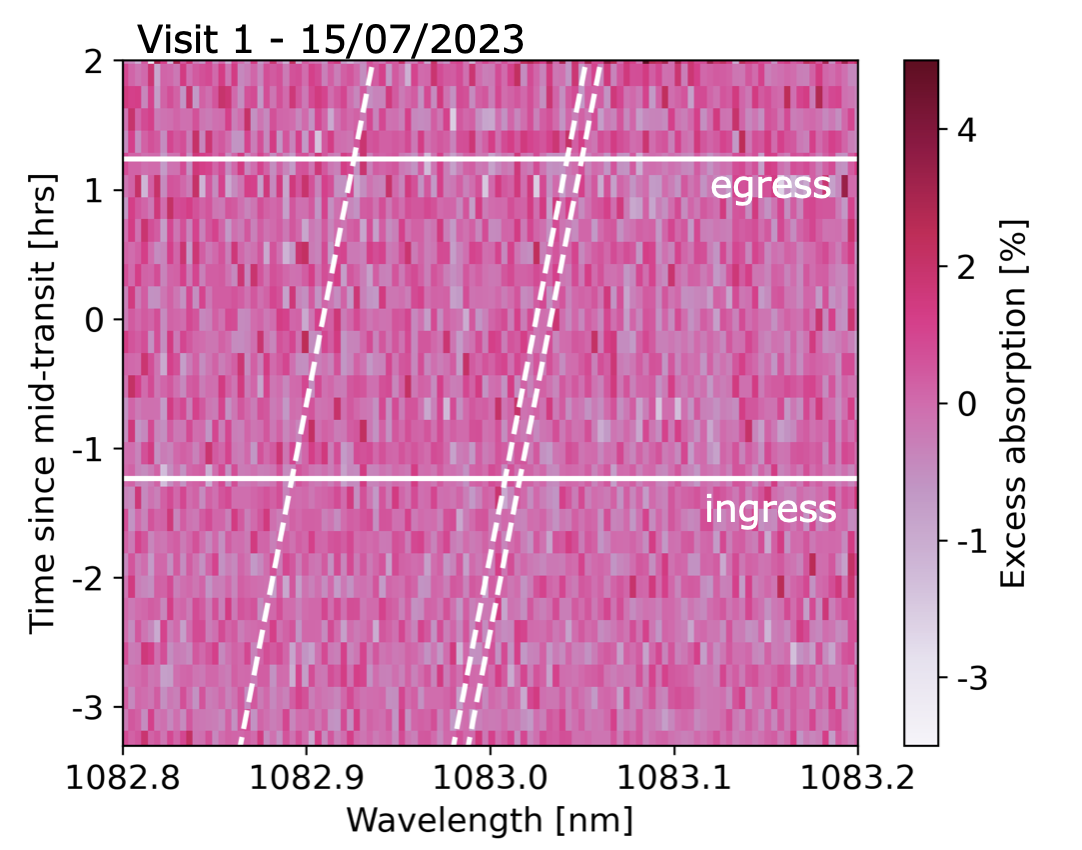}  
    \includegraphics[width=0.91\linewidth]{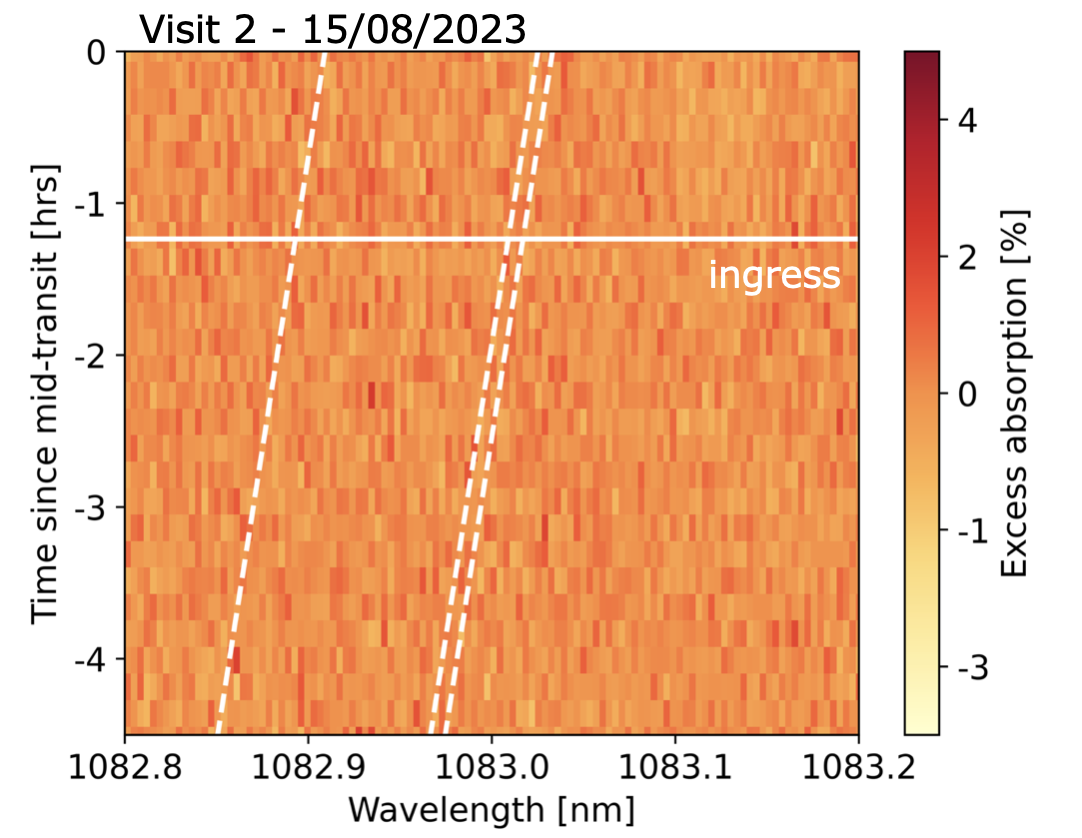}
    \caption{Spectral residuals as a function of time since mid-transit (hours) and wavelength (nm). Visit 1 is in pink and visit 2 is in orange. The solid horizontal lines represent ingress and egress. The slanted dashed lines mark the wavelengths of the metastable helium triplet of lines, Doppler shifted to the rest frame of the planet. The colorbar on the right represents the amount of excess absorption within the plot in percent.}
    \label{fig:helium_excess_abs}
\end{figure}

Fig~\ref{fig:helium_excess_abs} shows the spectral residuals over the helium infrared triplet as seen over both visits. Over-plotted are lines marking the expected radial velocity variation of the planet, and the times of ingress and egress. We detect no significant excess absorption or variability of the helium triplet from these spectral residuals.

% HIP 94235b has a radial velocity semi-amplitude ($K_p$) of 111 km/s. We calculated the phase-dependent RV in order to calculate the doppler shift for the metastable helium triplet of lines of liness in the stellar rest frame. The dashed white lines in Fig~\ref{fig:helium_excess_abs} trace the regions where we would expect to see a signal if we were to detect helium within the planet's atmosphere. As is evident by both visits in Figure~\ref{fig:helium_excess_abs}, we report a null-detection of escaping helium from the atmosphere of HIP94235b.

\begin{figure}[h!]
    \centering    \includegraphics[width=0.9\linewidth]{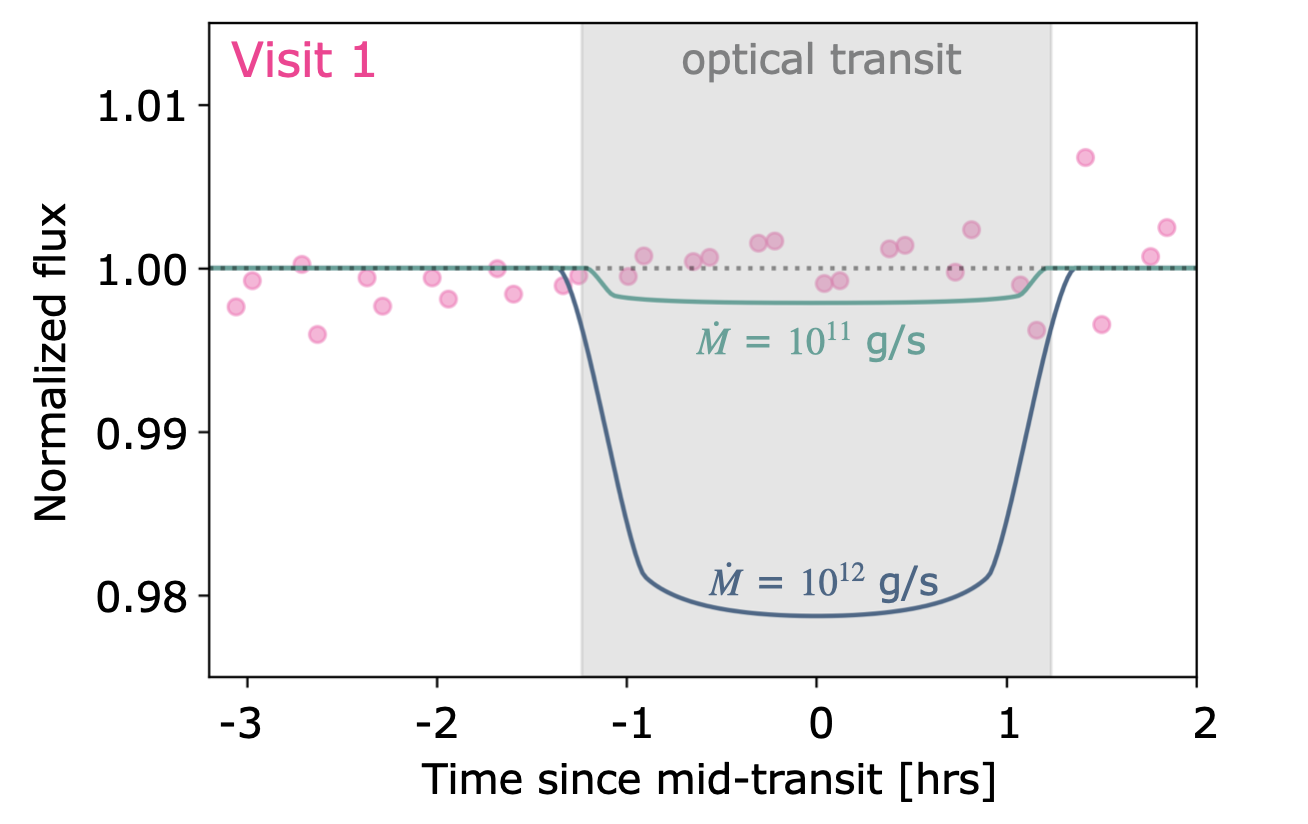}  
    \includegraphics[width=0.9\linewidth]{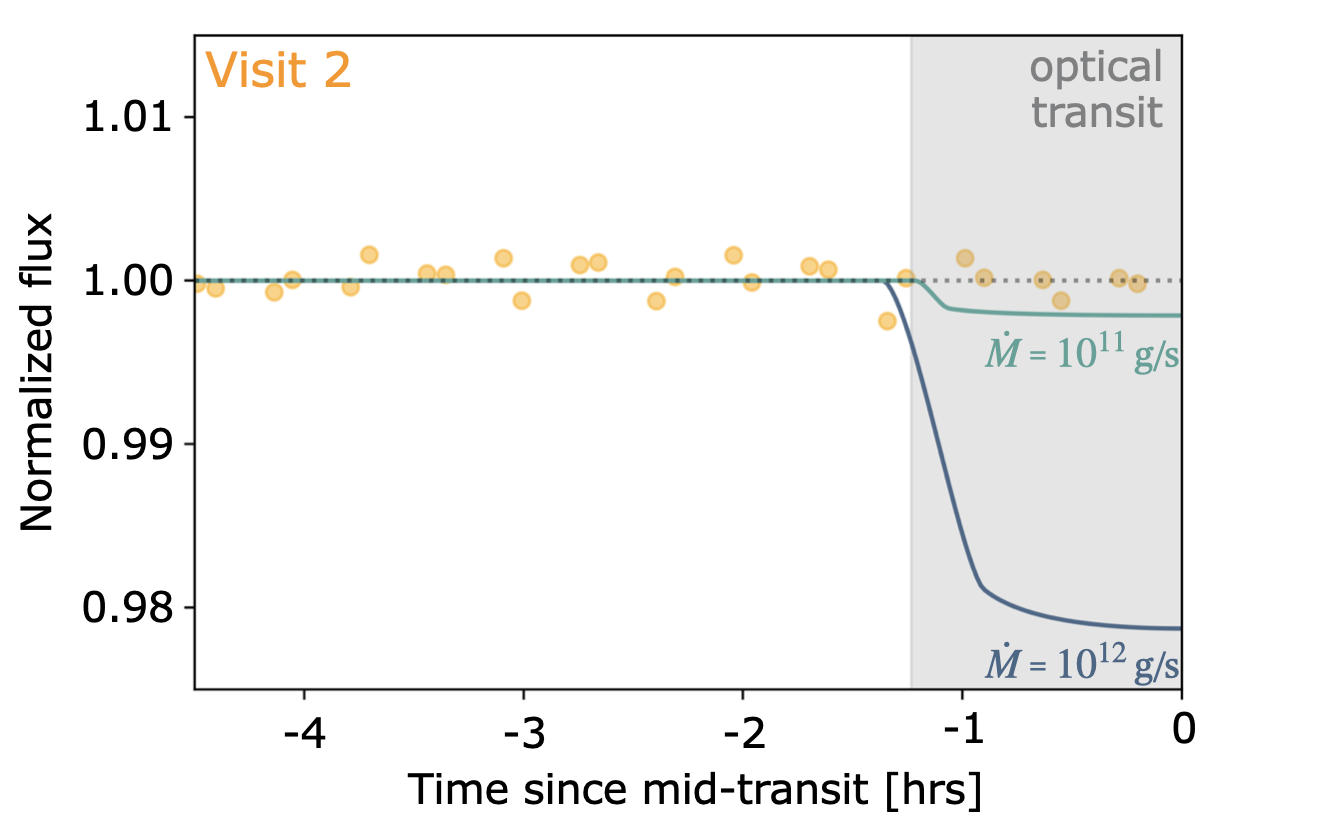}
    \caption{Lightcurves for visit 1 (top) and visit 2 (bottom) that trace along where we would expect the metastable helium triplet of lines to be present. The optical transit of HIP94235b is depicted by shaded grey regions. Lightcurves with transit depths based on different mass loss rates are shown in teal ($\dot{M}$=$10^{11}$gs$^{-1}$) and blue ($\dot{M}$=$10^{12}$gs$^{-1}$).}
    \label{fig:helium_lightcurves}
\end{figure}

%\subsection{Helium infrared triplet lightcurves and mass loss constraints}
Before computing the helium absorption light curve, we first shifted each spectrum to the planetary rest frame using it's radial velocity before combining them. Then, to quantify any flux variability over the helium lines, we compute the total flux over the helium infrared triplet over each exposure. We integrated within 0.15nm of the main helium triplet peak at 1083.33nm, as per \citet{Zhang_2023} to calculate the helium lightcurves, shown in Fig~\ref{fig:helium_lightcurves} for both visits. We detect no variability over the helium wavelength over the course of the planetary transits. 

To determine mass loss upper limits that can be established, we first fit for the shallowest transit depth that would be detectable in these observations. We model the transit as per \citet{2002ApJ...580L.171M} via the \textsc{batman} code \citet{2015PASP..127.1161K}. We allow $R_p/R_\star$ to be a free parameter, while holding the remaining transit parameters constant, including $a/R_\star$, inclination, and transit period and centroids as per the white light transit. We find we can place a $5\sigma$ upper limit of $R_p/R_\star < 0.03919$ ($<0.0015$ in transit depth) via the posterior distribution.

To convert this transit depth limit to a mass loss limit ($\dot{M}$), we make use of the \textsc{p-winds} Parker-winds model \citep{2022A&A...659A..62D}. \textsc{p-winds} is a 1-D wind model designed to simulate the interaction between the XUV flux of a host star and the escaping hydrogen and helium from the orbiting planet. The model depends on key system parameters, including stellar flux, planetary composition, and atmospheric structure, to predict mass loss rates. We make a number of assumptions in this model, the temperature of the upper planetary atmosphere is set to 6000 K, with the hydrogen number fraction assumed to be 0.9, and the remaining 10$\%$ allocated to helium. Adopting the same approach as \citet{Morrissey2024}, we scaled the XUV flux from the 100 Myr old G-type star EK Dra \citep{2005ApJ...622..680R} to estimate the incident flux for HIP94235b.

Example transit models with depths corresponding to reaonsable mass loss rates are over-plotted in Figure~\ref{fig:helium_lightcurves}. From these, we place a 5$\sigma$ upper limit constraint on the ongoing mass loss rate of HIP94235b as $10^{11}$g$s^{-1}$, or 0.53 M$_\oplus$/Gyr. Specifically, we find mass-loss rate constraints of 7.5$\times$10$^{10}$gs$^{-1}$ for visit 1, 3.9$\times$10$^{10}$gs$^{-1}$ for visit 2 and 3.9$\times$10$^{10}$gs$^{-1}$ from a joint fit to both visits. We report our final upper limit as $10^{11}$g$s^{-1}$, rounded up to one significant figure to reflect the uncertainties associated with our use of a simplified one-dimensional Parker wind model through our use of \emph{p-winds}, and to emphasis that our results are constrained on the order of magnitude level.\\

\subsection{Constraining planet mass loss evolution history}
We follow \citet{Morrissey2024} and constrain XUV driven photoevaporation tracks via the mass loss upper limits we determined from the helium null detection. We model the radius and envelope mass evolution of HIP94235b as per the energy limited approximation detailed in \citet{Owen_2017}, finding sets of models that will fit the current age and radius of the planet. We free the core mass, initial envelope fraction, and mass loss efficiency $\eta$ of the planet. 

\begin{figure*}
    \centering
    \includegraphics[width=0.9\linewidth]{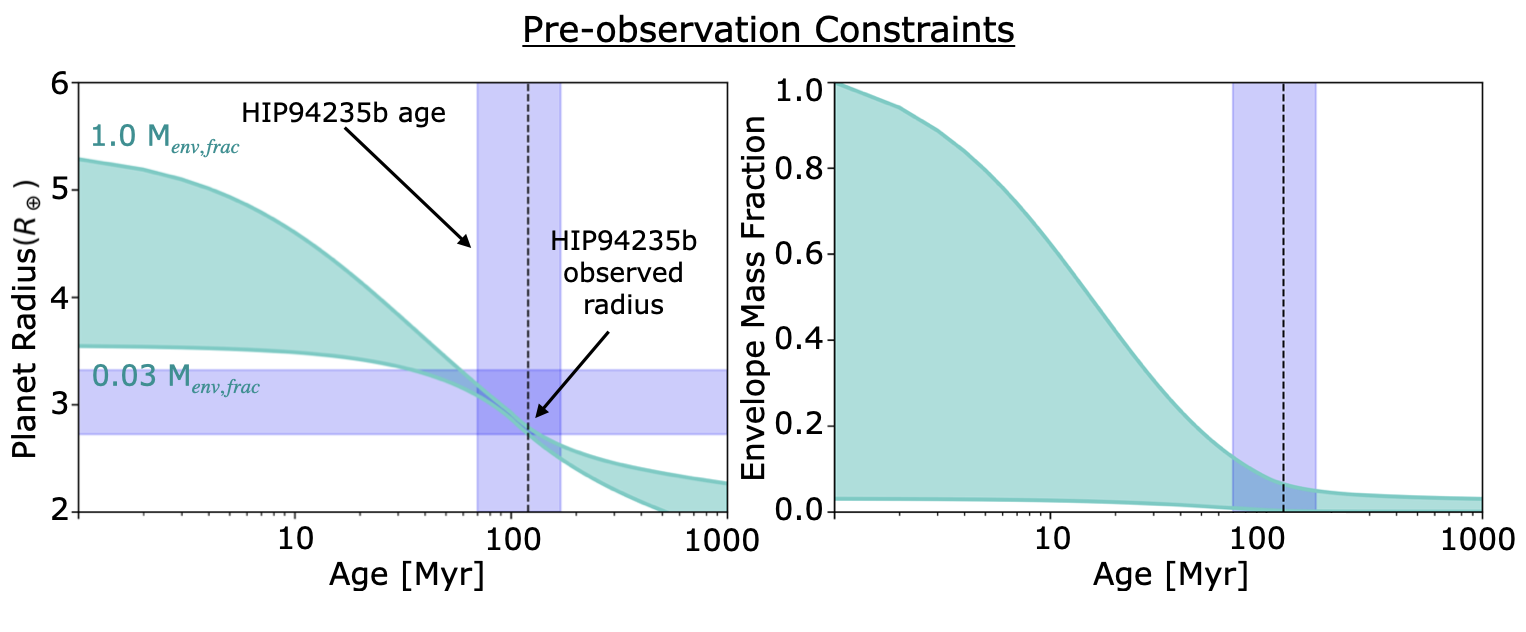}
    \includegraphics[width=0.9\linewidth]{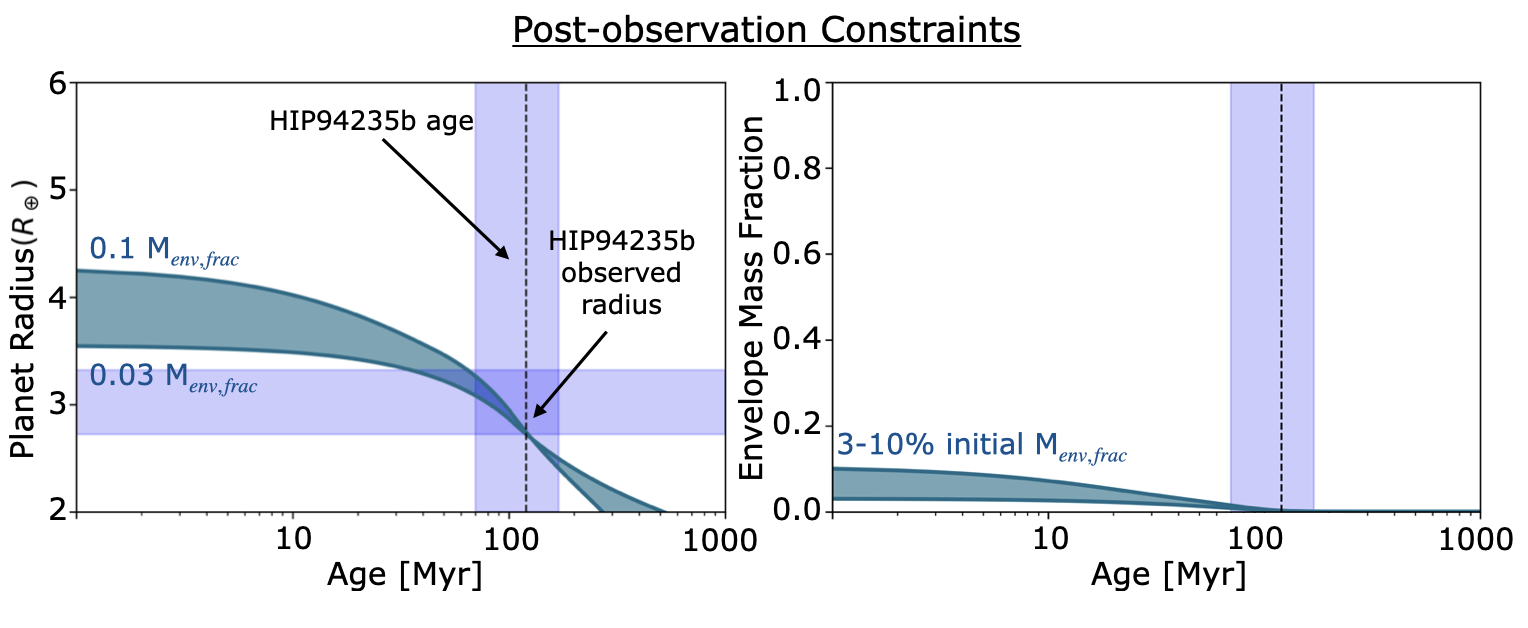}
    \caption{Evolution tracks for HIP94235b's radius (left) and envelope mass fraction (right) evolution over the first Gyr. The top panel shows the allowed tracks before these CRIRES+ results, with the bottom panel showing the observationally constrained tracks. HIP94235b has formed with a radius of 3.5-4.2 R$_\oplus$ and an initial envelope mass fraction of 3-10$\%$. The transparent blue bars in each figure show the error on the planet radius (R = 3.00$^{+0.32}_{-0.28}$R$_\oplus$) as the horizontal transparent region, and planet age (120$\pm$50 Myr) as the vertical transparent region.}
    \label{fig:evtracks}
\end{figure*}

Figure~\ref{fig:evtracks} shows the allowed set of models that can replicate the present day radius of HIP94235b. Without the helium mass loss constraints, the current conditions can be reproduced with a wide range of initial envelope mass fractions, ranging from 3\% to 100\% gas in composition, with corresponding core masses ranging from 2 to 10\,$M_\oplus$. 

Incorporating the observationally determined mass loss rate constraint of $\dot{M} < 10^{11}$gs$^{-1}$, these evolution models can be significantly constrained. In particular, the initial envelope mass fraction is expected to be 3-10\% of the mass of the planet. 

From these energy limited models, we find the low mass loss rate of HIP94235b is still consistent with having been born in a gas rich environment. \citet{2021MNRAS.503.1526R} showed the \emph{Kepler} small planets can be replicated by a population of `gas dwarfs' with initial gas envelopes that follow a $\beta$ distribution with a mean of $\sim 4\%$ in total planet mass. Significantly gas rich scenarios are likely ruled out for HIP94325b. Recent HST and JWST observations of the planets around the young systems of V1298 Tau \citep{2024NatAs...8..899B,2024A&A...692A.198B} and HIP67522b \citep{2024AJ....168..297T} showed they may consist with gas envelopes $\sim 40$\% in mass fraction. Such low density young planets are unlikely to be progenitors of HIP94235b. 

%Knowing an upper limit on the amount of mass being lost from HIP94235b at it's current age of 120Myr old, allows us to rewind the clock to understand how much gas HIP94235b formed with in it's primordial envelope. Through the new observational constraint on the amount of helium being lost, we can constrain the evolutionary tracks for HIP94235b over the first Gyr of it's lifetime. Figure~\ref{fig:evtracks} show the pre-constraints, allowed paths for both HIP94235b's planet radius on the left plot and envelope mass fraction (the amount of mass in the envelope relative to the core) on the right plot of the top row. The constrained evolutionary tracks for both parameters are depicted on the bottom row of plots, showing that HIP94235b formed with a radius of 3.5-4.2 $R_\oplus$ and an initial envelope mass fraction of 3-10 $\%$.\\
% \textcolor{orange}{note: check efficiency and parameters that went into models, what value was a efficiency of 0.1? - tightened efficiency constraint of 0.1 gives envelope around 4 percent}

\section{High resolution search for transmission spectroscopic features}

We also make use of the Y-band transit observations to search for additional spectral features in the atmosphere of HIP94235b. We cross correlate the CRIRES+ spectral residuals against planetary spectral templates to search for signatures of water vapor in the atmosphere of the planet. 

%Searching for atmospheric escape through the metastable helium triplet of lines can provide insights into the primordial atmosphere of HIP94235b and the rate at which it is evolving. Within the CRIRES+ Y-band, there is also an opportunity to search for species heavier than hydrogen or helium that may be present within this datset.\\ 

\subsection{Synthetic model spectra}\label{sec:synthetic_trans_model}
To generate model spectra for cross-correlation with our observations, we used the radiative transfer code \textsc{petitRADTRANS} \citep{Molliere_2019}. \textsc{petitRADTRANS} computes high resolution synthetic planetary transmission spectra via a line-by-line radiative transfer model.\\
We include absorption from water vapor, methane, carbon dioxide, and carbon monoxide in our line by line model. We also incorporated the effects of Rayleigh scattering via H$_2$ and He, alongside collision-induced absorption (CIA) from H$_2$-H$_2$ and H$_2$-He interactions.\\
The model atmospheres were computed over a pressure range of 10$^{-10}$ - 10$^{2}$ bar using 130 logarithmically spaced points.
The temperature-pressure (T-P) profile was computed using the analytical radiative equilibrium model from \citet{Guillot_2010}. Our chosen inputs for the infrared opacity ($\kappa_{IR}$), the ratio of optical to infrared opacity ($\gamma$), intrinsic temperature and equilibrium temperature were 0.01, 0.4, 200K and 1060K, respectively. Mixing ratios are computed via the grid from \textsc{EasyChem} \citep{2017A&A...600A..10M}, interpolated via the \textsc{interpol abundances} function in \textsc{petitRADTRANS}. 

\begin{figure*}
    \centering    
    \includegraphics[width=0.9\linewidth]{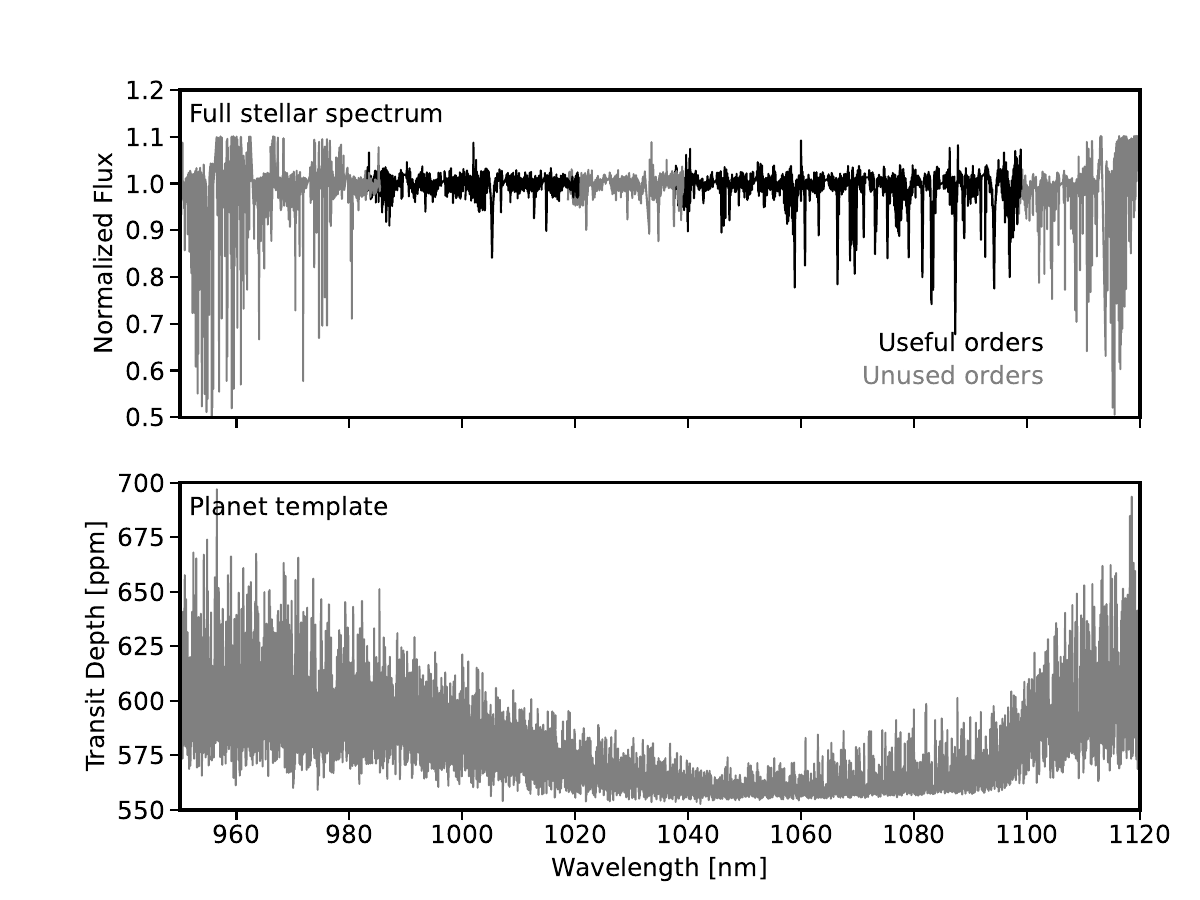}     
    \caption{\textbf{Top} The normalized and telluric corrected stellar spectrum over all orders are plotted, arranged in descending order from left to right, with order 9 on the far left and order 1 on the far right. Orders with observations included in our cross correlation analyses are marked in black, orders discarded in gray. \textbf{Bottom} The input synthetic transmission spectrum for a 3\,$R_\oplus$, 5\,$M_\oplus$ planet hosting a clear, solar metallicity atmosphere is plotted. All modeled lines within the spectral region observed come from water absorption.}
    \label{fig:prt_spectra}
\end{figure*}

We adopt system parameters for HIP94235b from \citet{Zhou_2022}. We assume a planet mass of $5\,M_\oplus$ for a low density sub-Neptune model with 2\% $H_2$ envelope at $3\,R_\oplus$ \citep{2019PNAS..116.9723Z}. We assume a cloudless atmosphere of solar metallicity, with a mean molecular weight of 2.3 atomic mass units, for our cross correlation analyses. The synthetic spectrum is shown in Figure~\ref{fig:prt_spectra}. 

% To explore the detectability of spectral features, we varied the planetary mass and radius values, generating a range of model spectra to determine the minimum planet size required to produce a detectable signal, as shown in Figure x. The surface gravity was then computed using Newton's law.\\
% The molecular mass fractions were assumed to be constant with pressure and set to X$_{H_2O}$ = 10$^{-3}$, X$_{CO}$ = 10$^{-2}$, X$_{CO_2}$ = 10$^{-5}$ and X$_{CH_4}$ = 10$^{-6}$. We assumed a mean molecular weight (MMW) of 2.33 atomic mass units (amu) throughout the atmosphere, consistent with a hydrogen-helium dominated atmosphere.\\

The model transmission spectra were computed using the \textsc{calc transm} function from \textsc{petitRADTRANS}, which solves the radiative transfer equation throughout a limb geometry to determine the effective transit radius as a function of wavelength. We defined the planetary radius and gravity at a pressure level of P$_0$ = 0.01 bar, providing the baseline for computing variations in spectral opacity along the planet's limb.\\
Given the resolving power of CRIRES+ ($R=100,000$), spectral features can be resolved down to a velocity difference of FWHM$\sim$3 km/s. To account for this instrument broadening, we defined a Gaussian function with a velocity dispersion of 3 km/s to match the resolving power. We then convolve the model spectrum with this Gaussian kernel to simulate the effects of instrumental broadening on our \textsc{petitRADTRANS} model spectra.\\
% \begin{figure*}[t]
%     \centering    \includegraphics[width=0.9\linewidth]{v1_masked_and_model.png} 
    
%     \caption{A sample observed spectra across all orders (1–9, from right to left) showing masked regions. CCD breaks and regions with large noise peaks have been manually masked. Orders 1, 8, and 9 are excluded due to significant noise. The \textsc{petitRADTRANS} model spectrum for a 3R$\oplus$, 11M$\oplus$ planet is shown in blue below the observed spectra, with shaded regions indicating the wavelength ranges where the observations and model spectra overlap.}
%     \label{fig:prt_spectra}
% \end{figure*}

\subsection{Stellar and telluric spectral removal}

Preparation of the observed spectra for cross correlations is performed similar to the procedure described in Section~\ref{sec:helium}. A telluric model was fitted for and removed from each observation via the \textsc{telfit} code as per Section~\ref{sec:telluric}. An example of one resulting stellar spectrum is shown in Figure~\ref{fig:prt_spectra}. As per the helium excess analysis, we extracted spectral residuals via the removal of a median spectrum. A master median combined spectrum was generated for each nodding position on each night using only the out of transit spectra, and removed from corresponding individual observations. This median master spectrum was generated at the stellar rest frame first. A second iteration of removals is performed at the telluric rest frame to remove any additional signals. The use of median spectral division over \textsc{SysRem} is motivated by the slow velocity variation of the planet over the course of the transit $(<10\,\mathrm{km\,s}^{-1})$. In section~\ref{sec:injection_recovery}, we demonstrate that it is difficult to preserve the planetary signal and efficiently remove stellar and telluric signals via \textsc{SysRem}. 

% SysRem was applied to both A and B nodding exposures separately to accurately remove systematics that differ within each (see Fig~\ref{fig:nodding} for differences between nodding A and B spectra). Once corrected for tellurics and stellar contamination, the A and B nodding exposures were re-joined together, where they were ordered by BJD before further analysis. The wavelength, phase and SysRem residuals (R) were extracted from the calibrated observations for cross-correlation with model spectra.\\
% \textcolor{orange}{add in other methods that arent SysRem here}\\

% \begin{figure}
%     \centering
%     \includegraphics[width=0.9\linewidth]{nodding_differences.png}
%     \caption{Spectra plotted from each .fits from visit 1 in subsequent order: nodding A (.00-.31), nodding B (.00-.31), with a vertical offset for plotting purposes. The change from nodding A spectra to nodding B spectra can be seen $\sim$ 4 on the y-axis.}
%     \label{fig:nodding}
% \end{figure}

Spectral regions near the detector edge, and regions heavily influenced by telluric absorption were manually masked from the dataset. Additionally, orders 1, 6, 8 and 9 were excluded from further analysis due to excessive noise (see Figure~\ref{fig:prt_spectra} for masked spectra).\\

\subsection{Cross-Correlation analysis}\label{sec:cc}

The spectral residuals from each epoch are cross correlated against our high resolution synthetic spectral template. A cosine apodization function was applied to the observed spectra to minimize edge effects. Cross correlation functions (CCFs) were computed over a velocity range of -300 to +300 $\mathrm{km\,s}^{-1}$ using a Doppler step of $3\,\mathrm{km\,s}^{-1}$. 

Cross correlation functions from select orders that are not telluric dominated (see Figure~\ref{fig:prt_spectra}) are median combined into one master cross correlation function per epoch. To remove common systematics that may still be present for all epochs, we also remove the common cross correlation function median combined over all epochs. This median correlation function is constructed for each night and each nodding position separately. 

\begin{figure}
    \centering
    \includegraphics[width=0.8\linewidth]{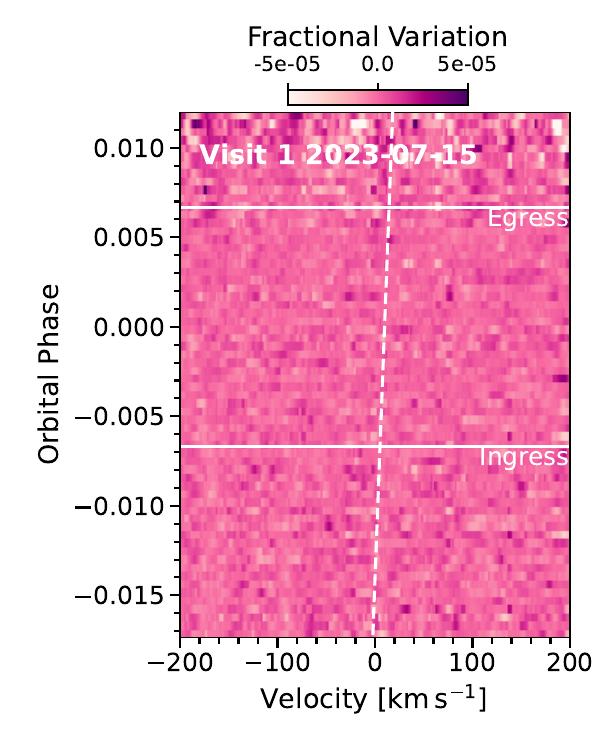}
    \includegraphics[width=0.8\linewidth]{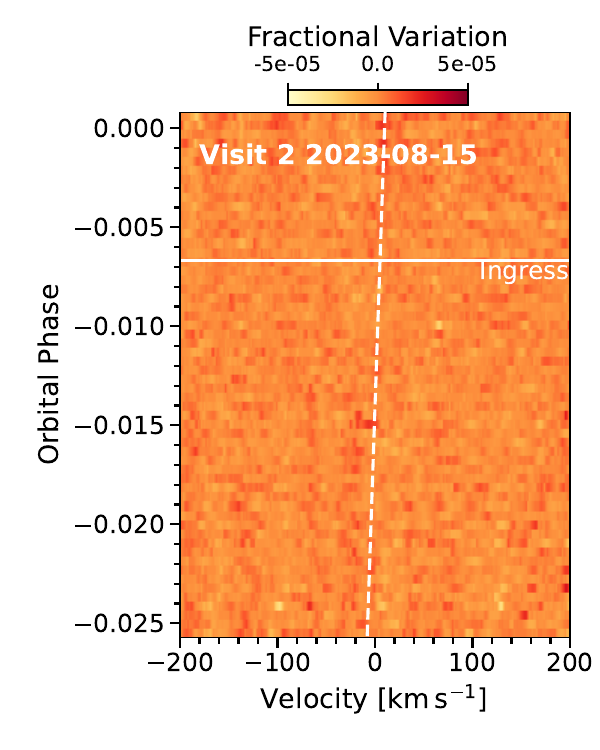}
    \caption{Cross correlation function variations over the two transit visits. Cross correlations are performed against a high resolution line by line spectral model of a $3\,R_\oplus$, $5\,M_\oplus$, solar metallicity atmosphere. The expected velocity variation of the planet, assuming a circular orbit, over the observations is marked by the dashed line. The times of ingress and egress are marked by the horizontal lines.}
    \label{fig:cc_visits}
\end{figure}

The temporal variations in the cross correlation function are presented in Figure~\ref{fig:cc_visits}. The expected radial velocity variation of HIP94235b over the course of the observations are marked, along with the expected times of ingress and egress. We find no significant planetary absorption over the Y-band from the two CRIRES+ transit observations. 

%\subsubsection{K$_p$ - V$_{sys}$ Map} This doesn't need to be a separate section

\begin{figure}
    \centering    
    \includegraphics[width=0.9\linewidth]{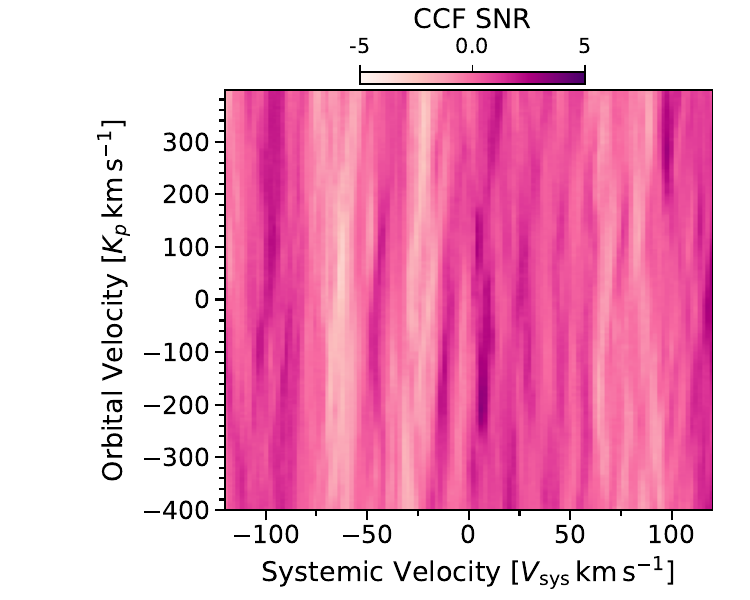}  
    \caption{Cross correlation peak signal to noise as a function of the assumed orbital and systemic velocities of the planet, averaged over both nights of observations. For a circular orbit, the expected velocities of the planetary system are $K_p = 111\,\mathrm{km\,s}^{-1}$ and $V_\mathrm{sys} = 9\,\mathrm{km\,s}^{-1}$. We find no evidence of excess absorption originating from the planet. }
    \label{fig:kp_vsys}
\end{figure}

If a planetary signal is present, we would expect the detection of excess absorption at the correct orbital and systemic velocity of HIP94235b. We map the signal to noise of our observations over a grid of orbital $K_p$ and systemic $V_\mathrm{sys}$ velocities. For each $K_p$ and $V_\mathrm{sys}$ grid point, we velocity shift and interpolate the cross correlation functions from each epoch to the rest frame of the planet, and median combine all in-transit observations over the two transits. To estimate the signal at each grid point, we follow previous examples \citep[e.g.][]{Brogi:2018,Gibson_2020} and measure the cross correlation function peak height over a central $\pm 2\,\mathrm{km\,s}^{-1}$ region. The noise in the combined cross correlation functions is characterized by the standard deviation in this peak height over velocity spaces surveyed at $|K_p| > 200\,\mathrm{km\,s}^{-1}$ and systemic $|V_\mathrm{sys}| > 200\,\mathrm{km\,s}^{-1}$. Figure~\ref{fig:kp_vsys} shows the signal to noise map over a range of reasonable orbital and systemic velocities. No significant absorption is identified at the velocities of the planetary system of $K_p = 111\,\mathrm{km\,s}^{-1}$ and $V_\mathrm{sys} = 9\,\mathrm{km\,s}^{-1}$ \citep{Zhou_2022,2021AA...649A...1G}. We note the eccentricity of the planet is not well constrained from photometry alone, and as such a detection at different $V_\mathrm{sys}$ and $K_p$ velocities would still have been acceptable as a planetary signal, with the effect of eccentricity on a detection shown in Figure 4 of \citet{Grasser_24} for GJ 436b. 

%The combined CCF's were phase-folded using the planetary ephemeris to seperate in-transit from out-of-transit exposures. This was to ensure that only in-transit planetary signals were considered while filtering out background noise. The strength of the retrieved CCF signal was found by computing the signal to noise ratio (S/N) within $\pm$2 km/s of the expected planetary velocity. The noise level was estimated using out of transit spectra as a reference baseline. To visualise the signal detection, a grid search was performed across a range of planetary radial velocity semi-amplitudes (K$_p$) and systemic velocities (V$_{sys}$). This technique identified the most statistically significant detection by showcasing the most coherant signals across the K$_p$ - V$_{sys}$ parameter space.\\

\subsection{Injection and recovery tests}\label{sec:injection_recovery}
To quantify the detectability of planetary signals in our observations and our analyses, we inject the synthetic planetary template into our observations and test for its recoverability. We adopt the synthetic template generated in Section~\ref{sec:synthetic_trans_model}, and inject it into the extracted observed spectra prior to telluric and stellar removal. The injection is performed at the predicted orbital radial velocity of the planet and at a systemic velocity of $-25\,\mathrm{km\,s}^{-1}$ to avoid overlapping with any potential intrinsic planetary signal.

\begin{figure}
    \centering    
    \includegraphics[width=0.9\linewidth]{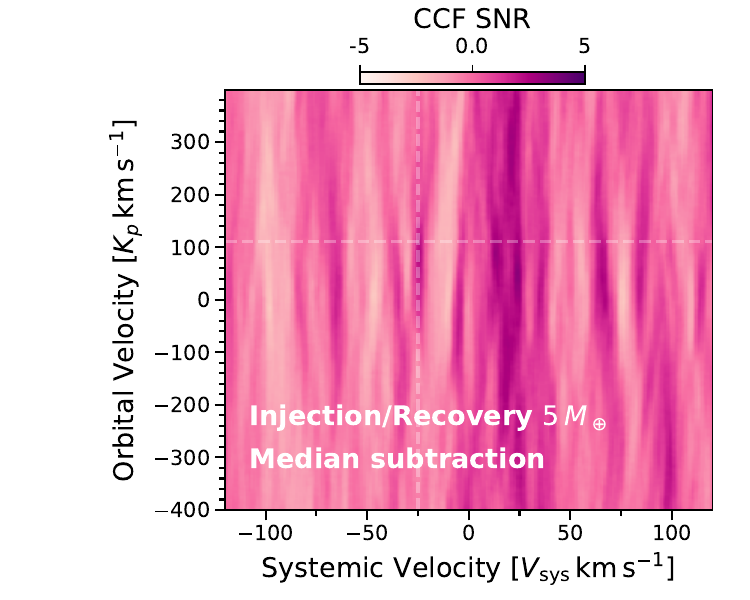}  
    \includegraphics[width=0.9\linewidth]{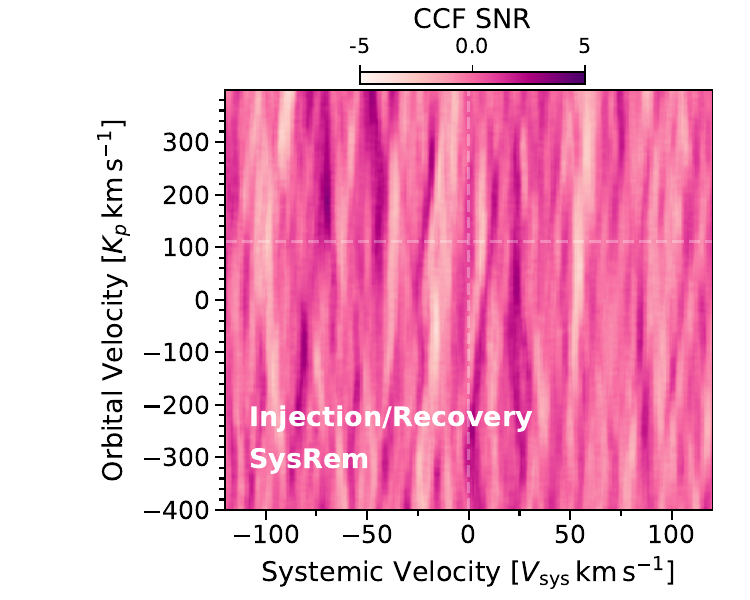}  
    \caption{Recovery of the injected transmission signal of a clear, solar metallicity planetary model. \textbf{Top} We find the planetary signal can only be recovered at the $2\sigma$ level when the stellar signals are subtracted via a simple removal of the median spectrum. \textbf{Bottom} \textsc{SysRem} removal of the stellar signal (over 4 iterations) removes the injected planetary signal, with no signals retrieved. The velocities of the injected signal are marked by the dashed lines. }
    \label{fig:kp_vsys_injectrecovery}
\end{figure}

The spectra are then cross correlated as per section~\ref{sec:cc}, and the signal detection is evaluated via the $K_p$-$v_\mathrm{sys}$ grid. We find that a $5\,M_\oplus$ solar metallicity cloudless planet model can be recovered when the datasets from both nights are combined at the $2\sigma$ level (Figure~\ref{fig:kp_vsys_injectrecovery}). We also found $10\times$ solar metallicity template, at $5\,M_\oplus$, can be retrieved at the $3.0\sigma$ level. We note that though a template of solar abundance was adopted, the only effective absorber in the Y-band is H$_2$O. Injection and recovery of other absorbers, including CO, CO$_2$, and CH$_4$, show the observations are not capable of providing meaningful constraints on their presence, unsurprising given the lack of significant absorption features in the observed wavelengths of our observations. 

We also performed the same analysis, but making use of the \textsc{SysRem} algorithm \citep{Tamuz_2005} to remove the stellar spectrum instead of median division. We found the injected planetary signal cannot be recovered if \textsc{SysRem} is adopted (see bottom plot of Figure~\ref{fig:kp_vsys_injectrecovery}). We attribute this to the low velocity variation of the planet during the transit, resulting in the removal of the planetary signal alongside the stellar signal via \textsc{SysRem}.

We also performed a series of injection and recoveries for a range of other atmosphere models, ranging between $M_p$ of 5-10$\,M_\oplus$ and metallicities of 1 - 500$\times$ solar. We find the only model that can be retrieved at the $3\sigma$ level is the $10\times$ solar metallicity $5\,M_\oplus$ template. Since no mass constraint is available for this planet due to the activity and rapid rotation of the target star, we can only conclude that a solar metallicity atmosphere is unlikely if the planet is of low density. The recovery significance from these simulations are presented in Table~\ref{tab:significance_table}.

\begin{table}
    \centering
    \caption{Retrieval significance for injection and recovery of transmission signal }
    \label{tab:significance_table}
    \begin{tabular}{ccc}
    & $5\,M_\oplus$ & $10\,M_\oplus$\\ 
    \hline\hline
    $1\times$ Solar & $2.1\sigma$ & $1.4\sigma$ \\
    $10\times$ Solar & $3.0\sigma$ & $2.1\sigma$ \\
    $100\times$ Solar & $2.4\sigma$ & $2.0\sigma$ \\    
    $500\times$ Solar & $1.8\sigma$ & $1.5\sigma$ \\ 
    \hline
    \end{tabular}
\end{table}

\section{Discussions and Conclusions}
\label{sec:conclusion}

We obtained Y-band observations of the 120\,Myr system HIP94235 via CRIRES+ on the VLT during the transits of its 7.7\,day period sub-Neptune. These observations were used to search for signatures of excess helium absorption that may be attributable to mass loss from the young planet. We also searched for additional absorption from the presence of water in the atmosphere of this sub-Neptune.

No excess helium absorption was detected for HIP94235b. The lack of mass loss detection is consistent with constraints placed via Lyman-$\alpha$ null detections of the same planet from \citet{Morrissey2024}. Adopting 1D Parker wind models from \citet{2022A&A...659A..62D}, we place a mass loss rate upper limit of $10^{11}\,\mathrm{g\,s}^{-1}$ for the planet. This mass loss rate limit is consistent with photoevaporation models that limit the initial envelope mass fraction to $<10\%$ of the total mass of the planet. Larger initial envelopes would have yielded stronger active mass loss rates at the current age of the planet. \citet{2021MNRAS.503.1526R} found similar gas envelope fractions can reproduce the close in super-Earth and sub-Neptune planet population from \emph{Kepler}. 

In contrast, larger low density young planets, such as HIP67522 b \citep{2024AJ....168..297T} and V1298 Tau b and c \citep{2024NatAs...8..899B,2024A&A...692A.198B}, have transmission spectra consistent with low density low metallicity atmospheres. The gas rich envelopes for these planets are expected to account for $\sim 40\%$ of the mass of the planet. If HIP 94235 b had progenitors similar to HIP67522 b and V1298 Tau b, we would have detected mass loss in our helium observations. 

We also searched for atmospheric water absorption in the transmission spectrum of HIP94235b. We report no detectable signatures from two transit visits. We show via injection and recovery exercises that select models of cloudless atmospheres are retrievable if present.

Ground based high resolution transmission spectroscopy has been attempted for few Neptune-sized planets. \citet{2024A&A...686A.127B} report detections of H$_2$O and NH$_3$ in the atmosphere of HAT-P-11b via GIANO-B observations spanning the wavelength range of $950-2450$\,nm on the Telescopio Nazionale Galileo. The atmosphere of GJ 436 b was probed by CRIRES+ over the wavelength range of 1490-1780\,nm, with a null detection of atmospheric signatures that is consistent with a high metallicity ($>300\times$ Solar) cloudy atmosphere, consistent with subsequent eclipse observations from JWST \citep{2025arXiv250217418M}. Similar high metallicity interpretations were inferred from CRIRES+ observations of the $2\,R_\oplus$ sub-Neptune GJ3090b \citep{2025arXiv250316608P}. JWST observations of other mature-aged sub-Neptunes have also found a predominance of such high metallicity atmospheres (e.g., TOI-836c, \citealt{2024AJ....168...77W}; GJ 9827 d, \citealt{2024arXiv241003527P}; GJ 1214 b, \citealt{2023ApJ...951...96G}). Such heavy atmospheres may still be the end product of photoevaporation in gas-rich planets \citep[e.g.,][]{2016ApJ...831..180C,2020ApJ...896...48M}, or they may be steam-based atmospheres in volatile-rich planets \citep[e.g.,][]{2020A&A...643L...1V,2021A&A...649L...5B}.

Within this context, we demonstrated that we are close to the point where ground-based observations can help rule out high metallicity atmospheres in scenarios such as HIP94235b. If the planet has such a high metallicity atmosphere, it would be similar to other small planets surveyed so far. The lack of active escape signatures and the high metallicity is consistent with the interpretation that either small planets are the end product of runaway mass loss, or that they are born with heavy, volatile rich envelopes. Differentiating between these scenarios will require observations of younger sub-Neptunes, for which density and mass loss rates between the two models should be substantially different \citep[e.g.][]{2025arXiv250317364R}.

\acknowledgements  
We respectfully acknowledge the traditional custodians of all lands throughout Australia, and recognise their continued cultural and spiritual connection to the land, waterways, cosmos, and community. We pay our deepest respects to all Elders, ancestors and descendants of the Giabal, Jarowair, and Kambuwal nations, upon whose lands this research was conducted.
GZ and AM thank the support of the ARC DECRA program DE210101893 and ARC Future Fellowship award FT230100517.
CH thanks the support of the ARC DECRA program DE200101840 and ARC Future Fellowship award FT240100016.

\facility{VLT,TESS}
\software{emcee \citep{2013PASP..125..306F}, batman \citep{2015PASP..127.1161K}, astropy \citep{2018AJ....156..123A}, PyAstronomy \citep{pya}, pycrires \citep{pycrires}}

% \appendix
% \section{Atmospheric retrievals for different assumed masses and metallicities}\label{sec:injectiongrid}
% Following the injection and retrieval exercise described in Section~\ref{sec:injection_recovery}, we performed a set of injection recoveries over a set of models covering planet masses of 5 and 10 $M_\oplus$ masses, and metallicities of 1, 10, and 100$\times$ solar. The resulting signal to noise of the recovery, parameterized by in the systemic and orbital semi-amplitude space, in shown in Figure X. 

% \begin{figure*}
%     \centering
%     \begin{tabular}{cc}
%     \includegraphics[width=0.3\textwidth]{vsys_krv_injectionrecovery_3Re5Me_1xsolar.pdf} & \includegraphics[width=0.3\textwidth]{vsys_krv_injectionrecovery_3Re10Me_1xsolar.pdf} \\
%     \includegraphics[width=0.3\textwidth]{vsys_krv_injectionrecovery_3Re5Me_10xsolar.pdf} & \includegraphics[width=0.3\textwidth]{vsys_krv_injectionrecovery_3Re10Me_10xsolar.pdf} \\
%     \includegraphics[width=0.3\textwidth]{vsys_krv_injectionrecovery_3Re5Me_100xsolar.pdf} & \includegraphics[width=0.3\textwidth]{vsys_krv_injectionrecovery_3Re10Me_100xsolar.pdf} \\
%     \end{tabular}
%     \caption{Injection and recovery grid, showing retrievals of injected models at a range of planet masses and assumed metallicities. }
%     \label{fig:injectionrecoverygrid}
% \end{figure*}

\bibliographystyle{apj}
\bibliography{refs}

\end{document}